\documentclass[prb,preprint,groupedaddress,showpacs]{revtex4}
\usepackage{epsfig}
\usepackage{graphics}
\usepackage{natbib}
\begin{document}
\newcommand{\beq}{\begin{equation}}
\newcommand{\eeq}{\end{equation}}
\bibliographystyle{apsrev}

\title{Controlling wave function localization in a multiple quantum well structure}

\author{Anjana Bagga and Anu Venugopalan}

\thanks{Corresponding author. Email:anu.venugopalan@gmail.com}

\affiliation{University School of Basic and Applied Sciences, GGS Indraprastha University, \\ Sector 16C, Dwarka,New Delhi-110075, India}

\begin{abstract}
The dynamics of a wave function describing a particle confined in a multiple quantum well potential is studied numerically. As a consequence of quantum mechanical tunneling, an initial wavefunction designed to be localized in one well can  localize in the others after a certain time and hop between wells at times which  depends on the height and width of the barriers separating the wells. This feature could find potential application in controlling electron transport in quantum well heterostructures.

\end{abstract}

\pacs{03.65.Ge, 42.50.Md, 73.20.Dx, 02.30.Mv}

\maketitle
Recently, there has been a lot of interest in studying the dynamics of wavefunctions in simple heterostructures. The dynamics of wavefunctions and the time scales involved can be controlled by physically accesible parameters. This sets a path for designing novel electronic devices out of them. The work described in this paper deals with wavefunction dynamics in a confined system modelled by a composite potential describing multiple  quantum wells. The possibility of quantum mechanical tunneling across the barriers separating wells allows an initial wavefunction designed to be localized in one well to localize in the others after a certain time which  depends on the height and width of the barriers separating the wells. This  dependence provides the possibility of developing a handle to control wave function localization which in turn could find potentially important applications in controlling electron transport in quantum well nanostructures. 

The dynamics of initially localized wavefunctions in confined system have often been shown to exhibit 'revivals'. This means that after evolving for a certain time when the probability distribution of the localized wavefunction might spread, the wavefunction relocalizes and acquires its original shape\cite{PS, Zoller, Averbukh, Bluhm}.  This behaviour is a consequence of the fact that in such systems the initial wavefunction is bound and its time evolution is governed  by a discrete eigenvalue spectrum\cite{Robinett}. The simplest class of systems for which  fractional and full revivals can be seen are those for which the energy spectrum goes as $n^{2}$ , e.g., the infinite square well potential and the rigid rotator\cite{Bluhm, Stroud, Robinett}. A number of studies have reported quantum revivals in a variety of quantum systems like that of Rydberg atom wavepackets, molecular wavepackets and  wave packets in semiconductor quantum wells and nanostructures\cite{Garraway,Warren, Marc, Stolow} etc. A lot of interest has also been generated in studying the quantum control of the motion of wavepackets in quantum well structures\cite{qcontrol}. Studies show that such systems can be experimentally designed and the  time scales for wave function dynamics and features like wave function revivals can be controlled by physically accesible parameters. Such studies are also driven by the motivation for  novel electronic devices fashioned out of heterostructures.
The simplest class of systems for which  fractional and full revivals can be seen are those for which the energy spectrum goes as $n^{2}$ , e.g., the infinite square well potential and the rigid rotator\cite{Bluhm, Stroud, Robinett}. A number of studies have reported quantum revivals in a variety of quantum systems like that of Rydberg atom wavepackets, molecular wavepackets and  wave packets in semiconductor quantum wells and nanostructures\cite{Garraway,Warren, Marc, Stolow} etc. The dynamics in a a double well structure has also been studied recently by modelling it with a Dirac $\delta$ function potential in the middle of the infinite square well potential\cite{Leo}.   On the experimental front,   coherent oscillations of an extended electronic  wavepacket in a GaAs/AlGaAS double quantum well structure  have been observed where the wavepacket had been created by ultrashot pulse excitation. The  sample modeling this potential was a $170 A^{0}$ GaAs quantum well followed by a $17 A^{0}$ barrier of $Al_{0.35}Ga_{0.65}As$\cite{Leo1}. A lot of interest has also been generated in studying the quantum control of the motion of wavepackets in quantum well structures\cite{qcontrol}. All these studies show that such systems can be experimentally designed and the  time scales for wave function dynamics and features like wave function revivals can be controlled by physically accesible parameters. Such studies are also driven by the motivation for  novel electronic devices fashioned out of heterostructures.

The simplest quantum system to illustrate the feature of quantum revivals is the one dimensional infinite square well potential where the time evolution can be beautifully illustrated because of the existence of exact analytical expressions for the energy eigenvalues and eigenstates. This system has been studied in great detail in the context of wave packet revivals\cite{Bluhm, Stroud, Robinett}. This familiar system has a potential
\begin{eqnarray}
V(x)&=&0; \; 0<x<L \\ \nonumber
V(x)&=&\infty ;\; x<0,\; x>L,
\end{eqnarray}
with eigenvalues and eigenfunctions
\begin{eqnarray}
E_{n}&=&\frac{\hbar^{2}\pi^{2}n^{2}}{2mL^{2}},\\
\phi_{n}(x)&=&\sqrt{\frac{2}{L}}sin{\frac{n\pi x}{L}},\;  n=1,2,3...,
\end{eqnarray}
where $L$ is the length of the well and $m$ is the mass of the particle. 
The time evolution of a particle in an initial state $\psi(x,0)$  can be written as:
\begin{equation}
\psi(x, t) = \sum_{n}c_{n}\phi_{n}(x)e^{−\frac{iE_{n}t}{\hbar}}.
\end{equation}
Here $n$ is the quantum number and $\phi_{n}(x)$ and $E_{n}$ are the energy eigenstates and corresponding eigenvalues for the system. The
coefficients $c_{n}s$ are given in terms of the initial wavefunction, as $ c_{n}=\langle\phi_{n}(x)|\psi(x, 0)\rangle$. The well known feature in the infinite well potential is that the  energy eigenvalues are quantized and are exactly quadratic in $n$. In many studies it has been shown that the time evolution of a wave function driven by such a spectrum has an important time scale called the 'revival time' which is contained in the coefficients of the Taylor series of the energy eigenvalue spectrum. In an infinite square well potential, this is the time in which an initial wavefunction regains its initial form. This 'revival time' for an infinite well is given by
\begin{equation}
T_{rev}=\frac{4mL^{2}}{\hbar \pi}.
\end{equation}
One can see that this time goes as $L^{2}$, where $L$ is the length of the well. The dependence on $L$ gives us a physically tunable parameter to control this time scale. It can be easily seen from Eq. (4) that in this system such  revivals are 'exact', i.e. $\psi(x,t+T_{rev})=\psi(x,t)$. At intermediate times the wavefunction spreads or shows partial or fractional reviavls. Fig. 1(a),1(b),1(c) and 1(d) shows probability distributions for the time evolved wavefunction which was initially localized, at different times. However, not all systems have a purely quadratic energy dependence like the infinite well and would, in general, contain higher order terms which could lead to more complicated and qualitatively different dynamics. For example, the more realistic case of a finite well has  been studied where the energy eigenvalue spectrum is not quadratic. In this system initial localized wavpackets have been found to undergo revivals and 'superrevivals'\cite{AV}.  Many of these features have an interesting physical and mathematical analog in the classical Talbot Effect in waveguides where a field of wavelength $\lambda$ propagating through a multimode planar waveguide of width $b$ would regenerate at multiples of a distance $L=2b^{2}/\lambda$\cite{talbot}. 

In this paper we numerically study wave function dynamics in a composite potential which represents a multiple quantum well structure.The numerical scheme used to find out the eigenvalues and eigenfunctions incorporates the tight binding Hamiltonian model\cite{tb}.  Our analysis allows us to add in wells, each well separated from the other by a finite barrier whose height and width can be changed. In particular we study how a wavefunction initially designed to be localized in one well evolves with time to localize in another well. The time scale for this to happen depends on physically accesible parameters like the height and width of the barriers and the dimensions of the system. To illlustrate our results, we look at a simple situation where the dimension of the system and the width of the barriers are fixed and the  localization is studied as a function of the barrier height. For our study we have looked at two scenarios: the potential profile with two wells and the potential profile with four wells [Figs.3 and 6]. Our analysis allows us to generalize the study to many wells. On solving the chosen system for its eigenvalues and eigenfunctions, the time evolution of an initial localized wavefunction is studied. In each case, the initial wavefunction can be designed by choosing the appropriate combination of the eigenstates that localize it in a chosen well. The time evolution features are revealed  by  the probability distributions of the time evolved initial state and are also reflected in the absolute square of the autocorrelation functions. The time evolved wave function is according to Eq.(4). The autocorrelation function is essentially the overlap of the time evolved  wavefunction with its initial state, i.e.,$\langle \psi(x,0)|\psi(x,\tau)\rangle$. For the one dimensional bound system considered here where the wavefunction is expanded in terms of the eigenfunctions $\phi_{n}$ with eigenvalues $E_{n}$, it can be easily checked that the autocorrelation function is given by
\begin{equation}
A(\tau)=\sum_{n}|c_{n}|^{2}e^{\frac{i \bar{E}_{n} \tau}{\hbar}}.
\end{equation}
In our study $\tau=t/T_{rev}$, i.e., the time is scaled in terms of the revival time, $T_{rev}=4mL^{2}/\pi$, of an infinite well potential of the same total length $L$ and $ \bar{E}_{n}=T_{rev}E_{n}$.

Our results reveal that a wavefunction initially localized in one well hops between wells with time. The time at which the localization in a particular well happens depends on the height and width of the barriers separating the wells. This dependence allows a useful handle to control the motion of the wavefunction by choosing the appropriate height and width to decide the time scales in which the localization occurs in a particular well. Consider the simplest case of a symmetric two-well confining potential (see Fig.3). For our analysis we choose the two-well potential profile with the following arbitrary parameters: Total lenghth=$100nm$ and barrier width$\sim 4.2nm$. For our analysis we keep the  barrier width fixed and study the dependence for two values of barreir height: $0.5eV$, and $0.6eV$. For such a system the time evolution of an initial wavefunction can be studied as described in Eq.(4). Since such a system does not have analytical solutions for the eigen values and eigen functions, we need to solve the Schrodinger equation numerically. The numerical solutions for the first two eigenfunctions of this systems are plotted in Fig.3. For our analysis we look at the simplest possible situation where we just deal with the first two eigenfunctions. We can construct a wavefunction that is localized in each of the wells by appropraitely choosing the values of the coefficients $c_{n}s$ that superpose the two eigenstates. 
For example, if the state is to be localized in the first well, it would be described by the superposition:
\begin{equation}
\psi_{1}(x, 0) = c^{1}_{1}\phi_{1}(x)+ c^{1}_{2}\phi_{2}(x).
\end{equation}
It can be easily verified that the choice of $c^{1}_{1} = c^{1}_{2}=\sqrt{.5}$ gives us the desired localization of the wavefunction in the first well.  Similarly the wavefunction giving a localization in the second well is given by the choice of coefficients $c^{1}_{1} = -c^{1}_{2}=\sqrt{.5}$. Fig.4 illustrates these wavefunctions. One can see that it is possible to design wavefunctions which is localized in any of the two wells with appropriate choices of the coefficients and the time evolution shows the "hopping" of the localized wavefunction between the wells. Once the initial wavefunction and the eigenvalues and eigenstates are known, the time evolution according to Eq.(4) gives us the wavefunctions at various times. The time evolution shows that the wavefunction periodically hops between the two wells. In order to ascertain at what time and in which well the wavefunction has localized, it is  convenient to look at the correlation functions for specific wells. This is done by looking at the overlaps of the time evolved wavefunction with states corresponding to the wavefunction being localized in a specific well, i.e., $|\langle\psi_{1}(x,0)|\psi_{k}(x, \tau)\rangle|^{2}$, and  $|\langle\psi_{2}(x,0)|\psi_{k}(x, \tau)\rangle|^{2}$ (k=1, or k=2). For example, Fig.5 looks at the correlation functions for the time evolved wavefunction $\psi_{1}(x, \tau)$ (k=1), which was initially localized in the first (left) well. Fig. 5(a) and 5(b) are plots of  $|\langle\psi_{1}(x,0)|\psi_{1}(x, \tau)\rangle|^{2}$, and $|\langle\psi_{2}(x,0)|\psi_{1}(x, \tau)\rangle|^{2}$ respectively. It is obvious that whenever each of these functions is peaked, there will be a localization in that specific well. To illustrate this dependence on the barrier height, Fig. 5 illustrates this feature for two different barrier heights, keeping the width fixed. The figure clearly shows the periodic hopping between the two wells and illustrates that this hopping is faster for smaller barrier heights. Note that the time here is scaled with respect to $T_{rev}$, the revival time for a single infinite well of the same length. (Fig. 2 shows the autocorrelation function for a single infinite well.)

Our numerical scheme for solving the Shr\"{o}dinger equation for such a composite potential and incorporating the time evolution  is quite general and can easily incorporate more wells. We extend this analysis to a four-well potential.  Fig.6 illustrates the symmetric four-well confining potential with the first four eigenfunctions. We keep all parameters for this the same as those for the two-well potential. Once again, one can see that it is possible to design wavefunctions which are localized in any of the four wells with appropriate choices of the coefficients and the time evolution shows the "hopping" of the localized wavefunction between the wells. The respective correlation functions allow us to determine the exact times at which this would happen. For example, if the initial state is to be localized in the first well, it would be described by the superposition:
\begin{equation}
\psi_{1}(x, 0) = c^{1}_{1}\phi_{1}(x)+ c^{1}_{2}\phi_{2}(x)+c^{1}_{3}\phi_{3}(x)+c^{1}_{4}\phi_{4}(x),
\end{equation}
with the choice of coefficients $c^{1}_{1} = c^{1}_{2}=c^{1}_{3} = c^{1}_{4}=\sqrt{.25}$.  Similarly, wavefunctions localized in the second, third and fourth wells can be constructed with the correct combination of the four eigenstates. Fig. 7 illustrates these various wavefunctions. Fig. 8 looks at the correlation function for the time evolved wavefunction $\psi_{1}(x, \tau)$ which was initially localized in the first well. Figs. 8(a), 8(b), 8(c) and 8(d) are plots of $|\langle\psi_{1}(x,0)|\psi_{1}(x, \tau)\rangle|^{2}$, $|\langle\psi_{2}(x,0)|\psi_{1}(x, \tau)\rangle|^{2}$,$|\langle\psi_{3}(x,0)|\psi_{1}(x, \tau)\rangle|^{2}$, and $|\langle\psi_{4}(x,0)|\psi_{1}(x, \tau)\rangle|^{2}$ respectively. It is obvious that whenever each of these functions is peaked, there will be a localization in that specific well. As mentioned before, our numerical scheme for solving the Shr\"{o}dinger equation for such a composite potential and incorporating the time evolution  is quite general and can easily incorporate more wells. In Fig. 9 we show the symmetric six-well potential with its first six eigenfunctions. Table 1 shows various combinations of the six coefficients in the superposition for localizing a wavefunction in a chosen well. With any of these choices as initial state, the time evolution shows the "hopping" of the localized wavefunction between the wells. The respective correlation functions allows us to determine the exact times at which this would happen and the choice of the barrier height can be used to control the times at which localization in a particular well takes place. This can be implemented in physical systems like quantum well nanostructures where the barrier heights can be externally controlled.

In conclusion, we have numerically studied the dynamics of an initially localized wavefunction describing a particle confined in a multiple quantum well potential with a perspective of controlling its dynamics. The numerical scheme can easily be generalized to include any number of wells. We have looked at two simple scenarios of two wells and four wells to illustrate the dynamics and indicated how this can easily extend to six wells.  Our scheme allows us to design wavefunctions which are initially localized in a chosen well and look at their localization in various wells with time. We see that as a consequence of quantum mechanical tunneling, an initial wavefunction designed to be localized in one well can  localize in the others after a certain time which  depends on the height and width of the barriers separating the wells, giving us an important handle to control the dynamics. This  interesting feature arising out of the dynamics driven by Schr\"{o}dinger evolution in a simple confined quantum system could find potential application in controlling electron transport in quantum well nanostructures.

\section*{Acknowledgments}
AV acknowledges financial support from the Department of Science and Technology, Government of India.

\pagebreak

\begin{flushleft}
\begin{table}
\caption{\bf \large Coefficients for wavefunctions localized in various wells in the six-well potential. (Each coefficient has a multiplicative factor of $\frac{1}{\sqrt{6}}$)}
\begin{tabular}{|l|c|c|c|c|c|c|}
\hline
Well & $c_{1}$ & $c_{2}$ & $c_{3}$ & $c_{4}$ & $c_{5}$ & $c_{6}$ \\ \hline
First & 1 & 1 & 1 & 1 & 1 & 1 \\ \hline
Second & 1 & 1 & 1 & -1 & -1 & -1 \\ \hline
Third & 1 & 1 & -1 & -1 & 1 & 1 \\ \hline
Fourth & 1 & -1 & -1 & 1 & 1 & -1 \\ \hline
Fifth & 1 & -1 & 1 & 1 &-1 & 1 \\ \hline
Sixth & 1 & -1 & 1 & -1 & 1 & -1 \\ \hline
\hline
\end{tabular}
\end{table}
\end{flushleft}

\begin{figure}
\hspace*{-1.80in}
\epsfig{file=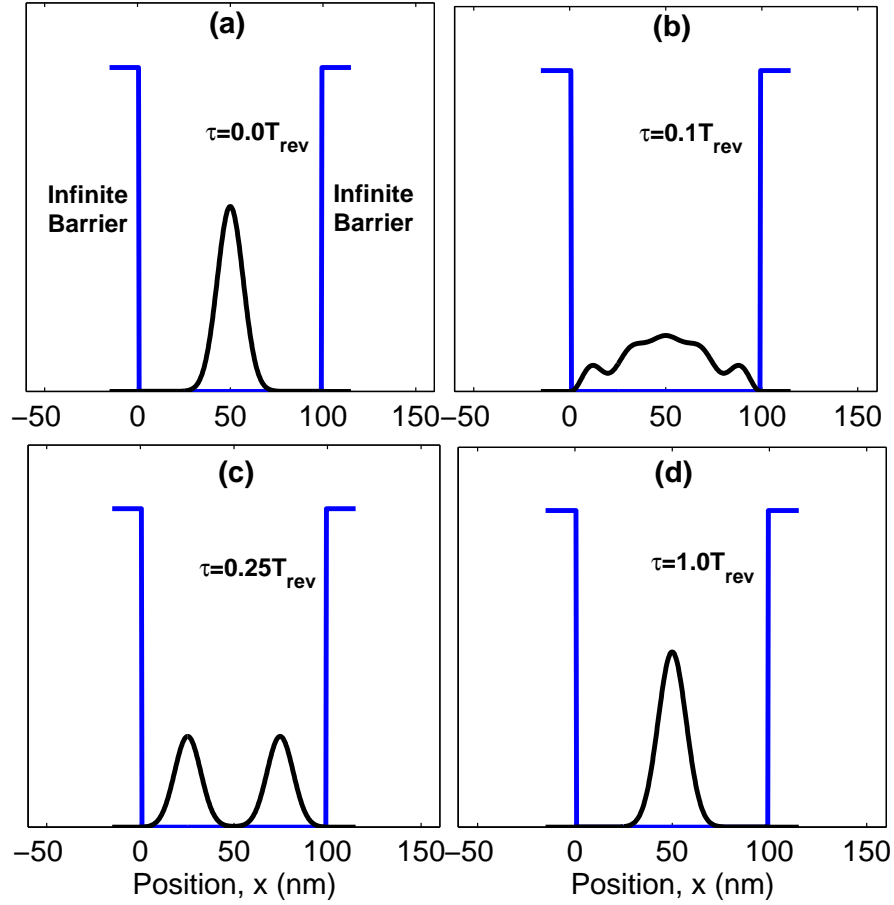, width=25cm} \\
\vspace{-0.3in}
\caption{\bf \large Initially localized wavfunction in the single infinite well. Probability distributions for (a) $(\tau=0)$ (b) $\tau=\frac{1}{10}T_{rev}$ (c) $\tau=\frac{1}{4}T_{rev}$; (d)$\tau=T_{rev}$ }
\end{figure}

\begin{figure}
\hspace*{-1.50in}
\epsfig{file=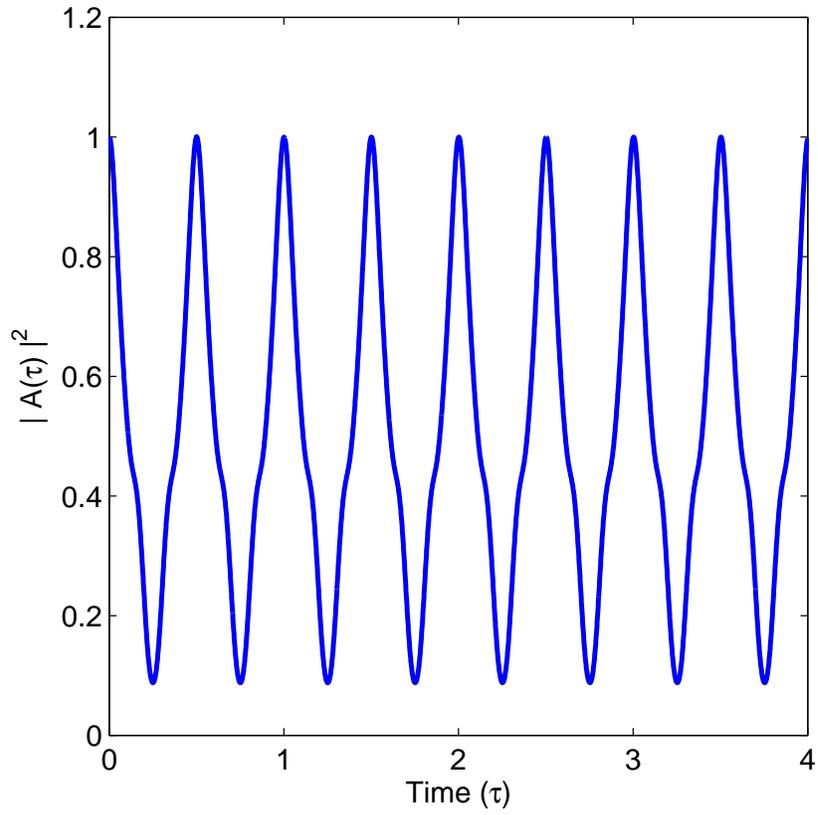, width=25cm}
\vspace{-1.5in}
\caption{\bf \large Square of the autocorrelation function $A(\tau)=\langle\psi(x,0)|\psi(x,\tau)\rangle$ for the single infinite well potential}
\end{figure}

\begin{figure}
\begin{flushleft}
\hspace*{-2.0in}
\epsfig{file= 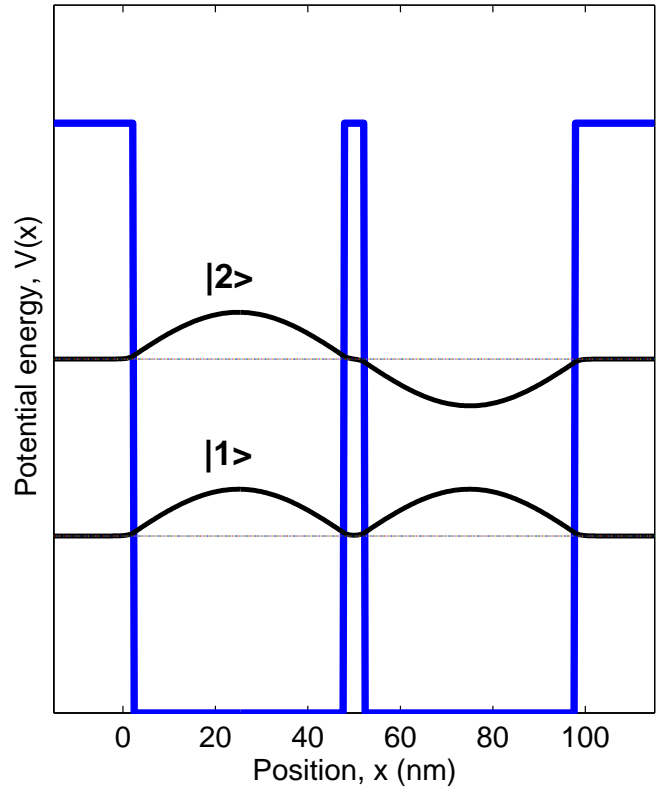, width=25cm}
\vspace{-1.0in}
\caption{\bf \large Two-well potential with the first two eigenfunctions}
\end{flushleft}
\end{figure}

\begin{figure}
\hspace*{-2.0in}
\epsfig{file= 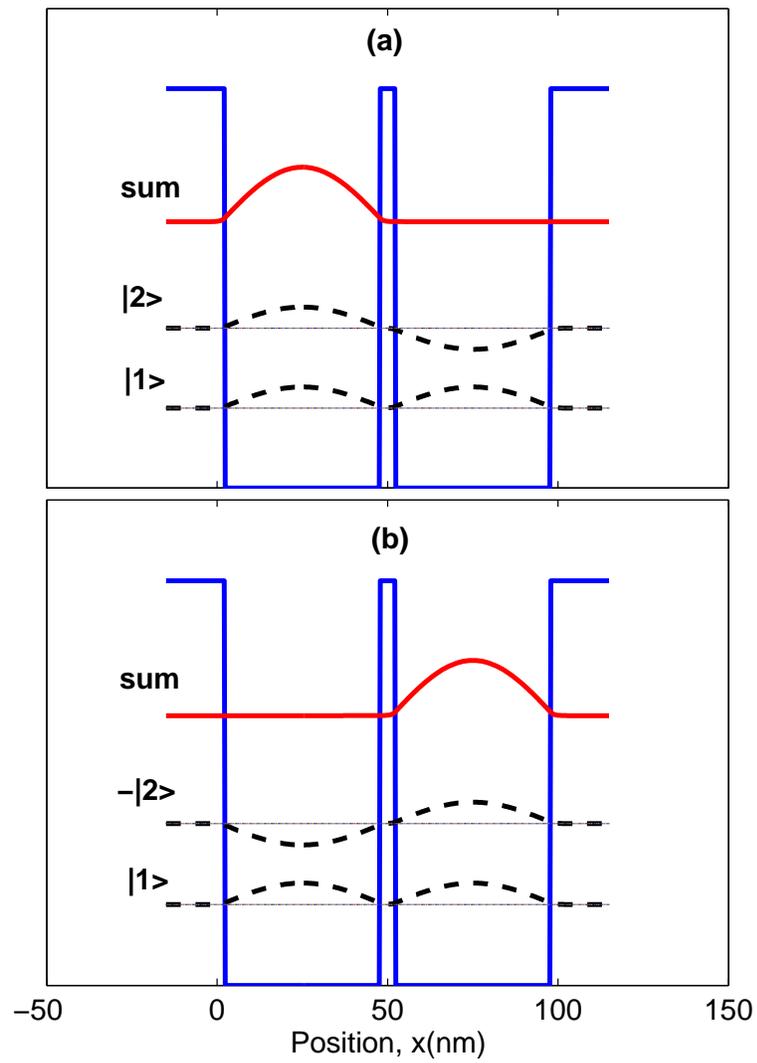, width=25cm}
\caption{\bf \large Two-well potential with wavefunctions designed to be localized in each well}
\end{figure}

\begin{figure}
\hspace*{-2.0in}
\epsfig{file= 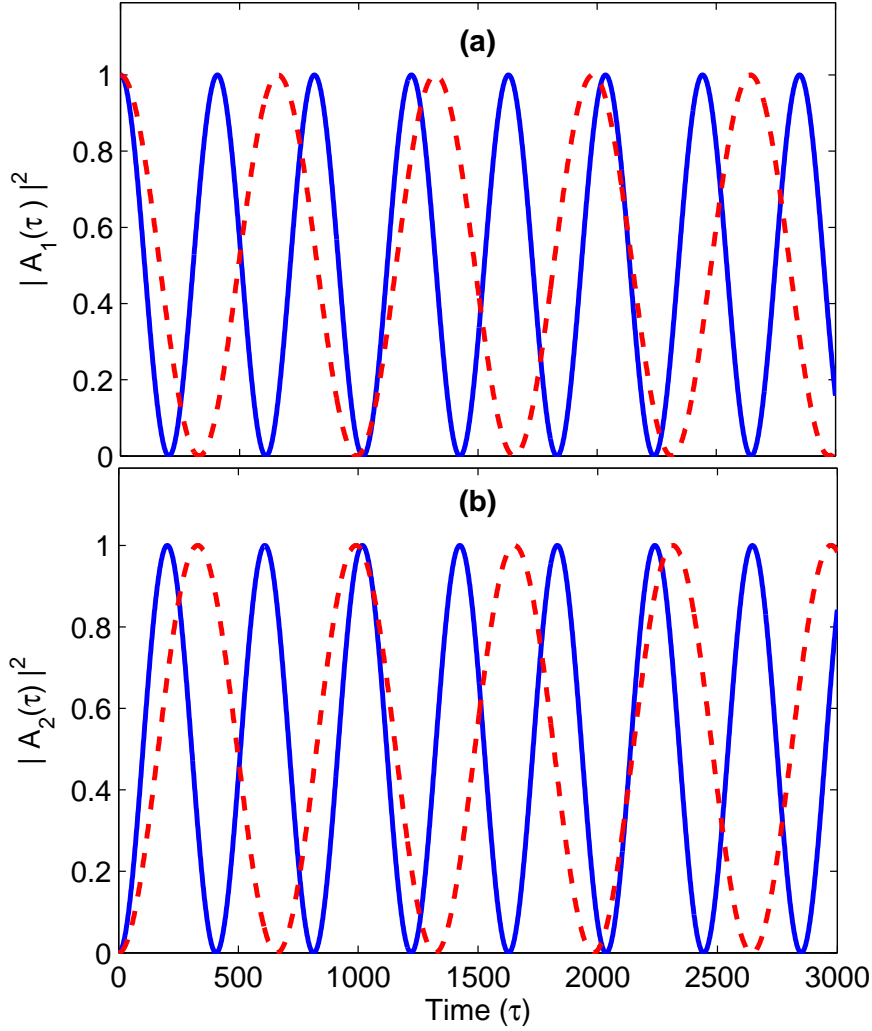, width=25cm}
\vspace{-0.5in}
\caption{\bf \large Square of the correlation functions for two-well potential with barrier heights =$0.5eV$ (solid line) and $0.6eV$ (dashed line). Peaks indicate localization in that particular well. (a) $A_{1}(\tau)=\langle\psi_{1}(x,0)|\psi_{1}(x, \tau)\rangle$, (b) $A_{2}(\tau)=\langle\psi_{2}(x,0)|\psi_{1}(x, \tau)\rangle$}. 
\end{figure}

\begin{figure}
\hspace*{-2.0in}
\epsfig{file= 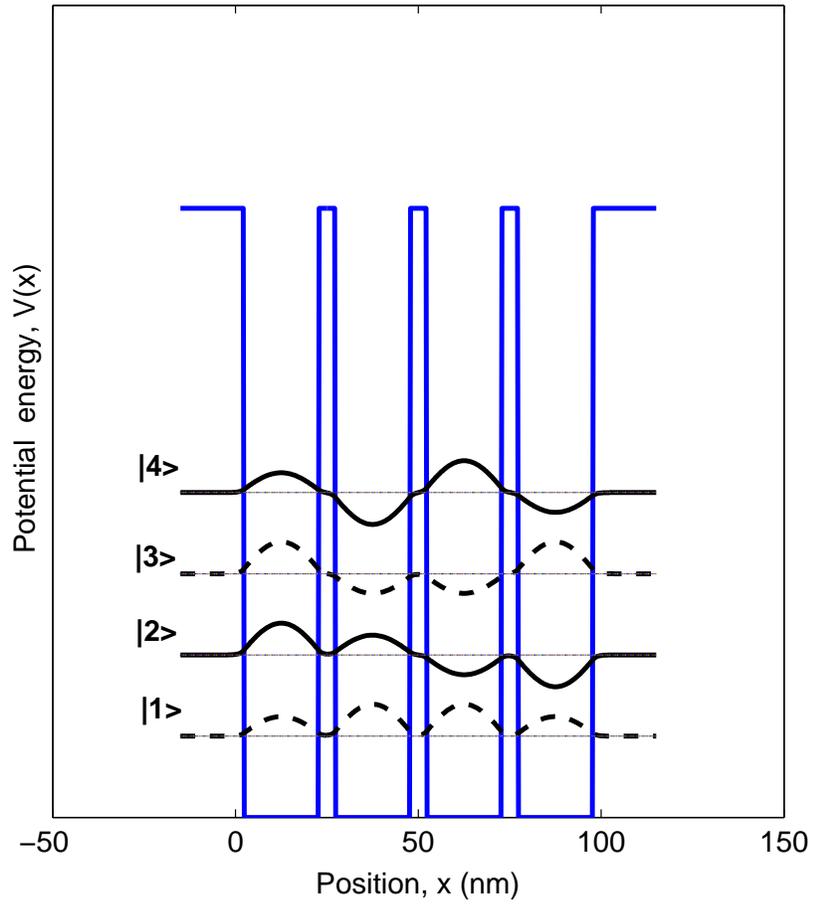, width=25cm}
\vspace{-0.5in}
\caption{\bf \large Four-well potential with first four eigenfunctions}
\end{figure}

\begin{figure}
\hspace*{-2.0in}
\epsfig{file= 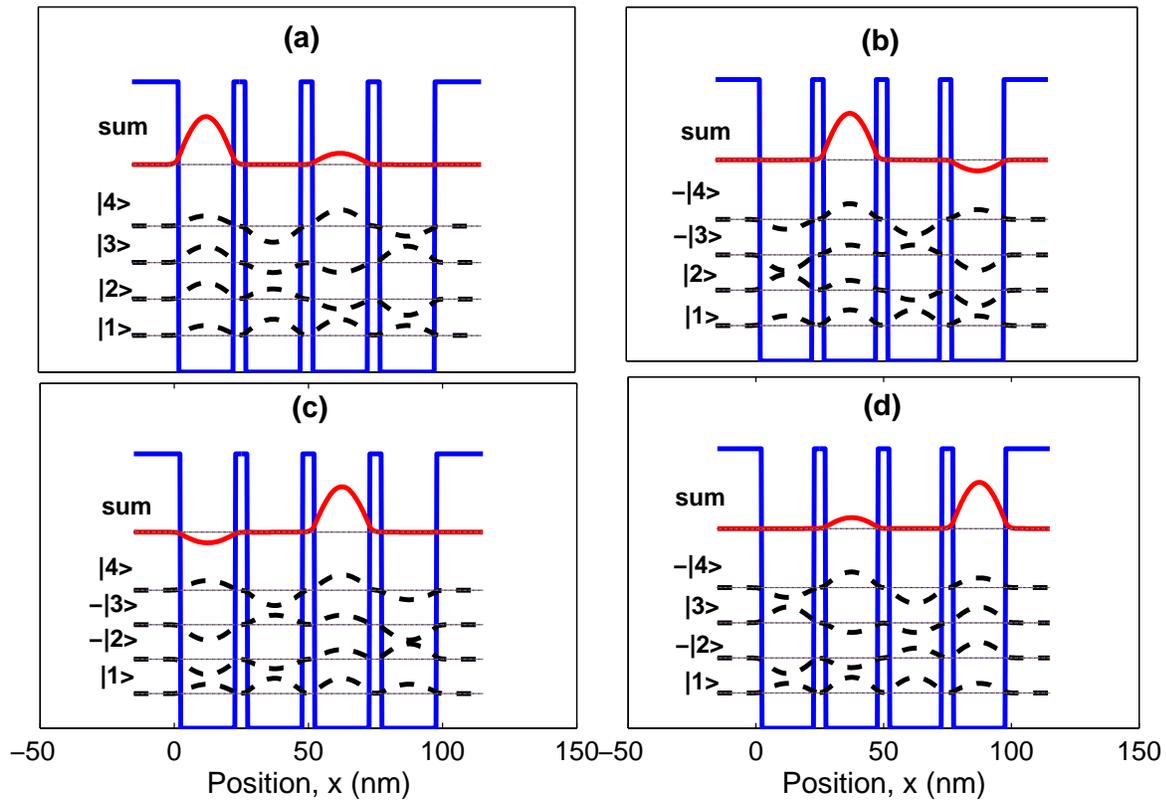, width=25cm}
\vspace{-0.5in}
\caption{\bf \large Four-well potential with wavefunctions designed to be localized in each well. Note that the wavfunction gets localized in a particular well when all amplitudes corresponding to the four eigenfunctions (with a chosen phase factor) in that well add up constructively}
\end{figure}

\begin{figure}
\hspace*{-2.0in}
\epsfig{file=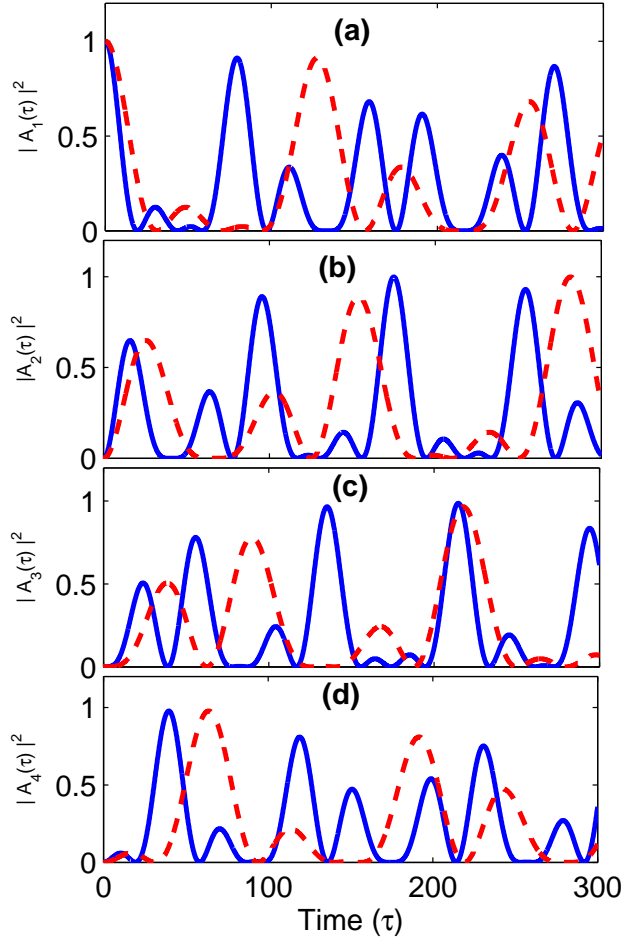, width=25cm}
\vspace{-0.5in}
\caption{\bf \large Square of the correlation functions for four-well potential with barrier heights =$0.5eV$ (solid line) and $0.6eV$ (dashed line)  Peaks indicate localization in that particular well.(a) $A_{1}(\tau)=\langle\psi_{1}(x,0)|\psi_{1}(x, \tau)\rangle$, (b) $A_{2}(\tau)=\langle\psi_{2}(x,0)|\psi_{1}(x, \tau)\rangle$, (c) $A_{3}(\tau)=\langle\psi_{3}(x,0)|\psi_{1}(x, \tau)\rangle$, (d) $A_{4}(\tau)=\langle\psi_{4}(x,0)|\psi_{1}(x, \tau)\rangle$}.
\end{figure}

\begin{figure}
\hspace*{-2.0in}
\epsfig{file=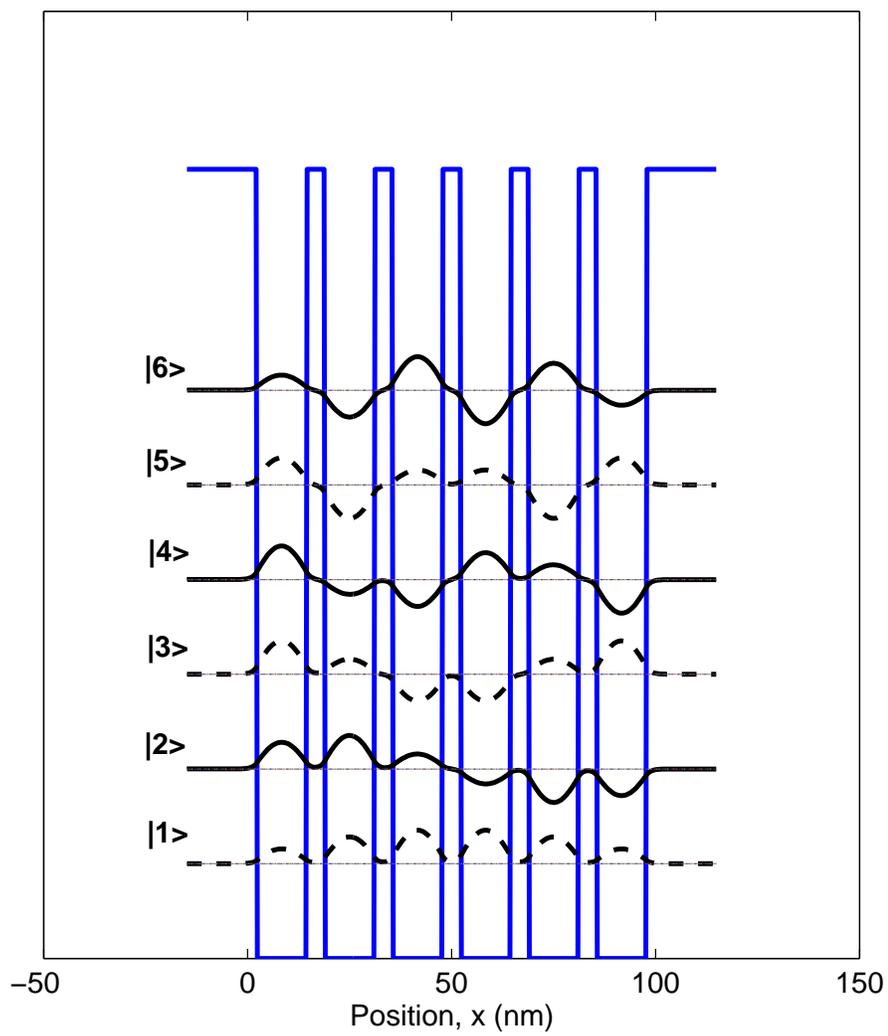, width=25cm}
\vspace{-0.5in}
\caption{\bf \large Six-well potential with first six eigenfunctions}
\end{figure}

\end{document}